\documentclass[%
 reprint,
amsmath,amssymb,
aps,
]{revtex4-2}

\usepackage{arydshln}
\usepackage{multirow}

\usepackage{graphicx}

\usepackage{dcolumn}
\usepackage{bm}

\usepackage[colorlinks,linkcolor=blue,citecolor=blue,urlcolor=blue]{hyperref}

\usepackage{booktabs}

\usepackage{textcomp}

\begin{document}

\preprint{}

\title{Density-wave order enhances the phonon thermal Hall effect in a trilayer nickelate}

\author{Qiaochao Xiang$^{1}$, Enkang Zhang$^{2}$, Xiaokang Li$^{1,*}$, Xiaodong Guo$^{1}$, Mengfei Zhu$^{1}$, Jun Zhao$^{2,*}$, Guang-Ming Zhang$^{3,4,*}$, Liang Li$^{1,*}$ and Zengwei Zhu$^{1,*}$}

\affiliation{
(1) Wuhan National High Magnetic Field Center and School of Physics, Huazhong University of Science and Technology, 430074 Wuhan, China\\ 
(2) State Key Laboratory of Surface Physics and Department of Physics, Fudan University, Shanghai 200433, China\\ 
(3) State Key Laboratory of Quantum Functional Materials, School of Physical Science and Technology, ShanghaiTech University, Shanghai 201210, China\\ 
(4) Department of Physics, Tsinghua University, Beijing 100084, China\\
}
\date{\today}

\begin{abstract}
Ruddlesden--Popper nickelates have emerged as a promising platform for high-temperature superconductivity, yet the role of lattice degrees of freedom in their correlated normal state remains largely unexplored. Here, we report the observation of a finite phonon thermal Hall effect in the trilayer nickelate La$_4$Ni$_3$O$_{10}$ at ambient pressure. Remarkably, the thermal Hall response is strongly enhanced below the density-wave transition at $T^*\approx140$ K, exhibiting two distinct plateaus in the thermal Hall resistivity. The characteristic energy scale extracted from the thermal Hall response ($\sim4.1$ meV) closely matches the magnon--phonon crossing span energy ($\sim3.2$ meV), pointing to magnon--phonon hybridization as the primary mechanism enhancing the thermal Hall effect. These results provide new insight into the interplay between lattice and spin excitations in nickelates, with implications for understanding both their superconductivity and the multiple possible origins of insulating thermal Hall signals.
\end{abstract}

\maketitle

The recent discovery of high-temperature superconductivity in nickelates has opened a new avenue for unconventional superconductivity\cite{li2019superconductivity,Pan2022,wang2022pressure,ding2023critical,sun2023signatures,zhang2024high,wang2024pressure,zhu2024superconductivity,shi2025absence,li2026bulk,li2024signature,li2025crystal,li2024structural,peng2026nearly,labollita2024electronic,luo2023bilayer,liu2023s,yang2023interlayer}. Following the discovery of superconductivity in infinite-layer nickelate films\cite{li2019superconductivity,Pan2022,wang2022pressure}, the bilayer Ruddlesden--Popper (RP) nickelate La$_3$Ni$_2$O$_7$ was found to exhibit superconductivity with $T_c$ approaching 80 K under high pressure\cite{sun2023signatures}, together with strange-metal behavior reminiscent of the cuprates\cite{zhang2024high,wang2024pressure}. These observations have stimulated extensive debate regarding the pairing mechanism and the role of magnetic correlations\cite{luo2023bilayer,liu2023s,yang2023interlayer}. The trilayer nickelate La$_4$Ni$_3$O$_{10}$ further enriches this family by exhibiting pressure-induced superconductivity with a maximum $T_c\approx30$ K\cite{zhu2024superconductivity,li2024signature,li2025crystal,peng2026nearly,li2024structural}. At ambient pressure, its normal state is characterized by a pronounced density-wave (DW) transition and intriguing interlayer coupling effects\cite{shi2025absence,zhu2024superconductivity,li2024signature,li2025crystal,peng2026nearly,li2024structural,labollita2024electronic}. However, the microscopic origin of the DW order---whether dominated by spin-density-wave (SDW) or charge-density-wave (CDW) correlations---and its coupling to the lattice remain unresolved\cite{Chan2026spin,Rustem2026,Chen2026SDW,keimer2015,comin2016,letacon2014}. More broadly, the relationship between antiferromagnetism and superconductivity in unconventional superconductors is known to involve both competition and cooperation rather than simple antagonism\cite{tranquada1995,kivelson2003}. While long-range antiferromagnetic order is typically suppressed by doping or pressure before superconductivity emerges\cite{keimer2015,kastner1998}, short-range dynamic spin fluctuations inherited from the antiferromagnetic background are widely believed to provide the pairing glue for Cooper pairs\cite{scalapino1995,fong1999}.

\begin{figure*}[ht]
\centering
\includegraphics[width=1\linewidth]{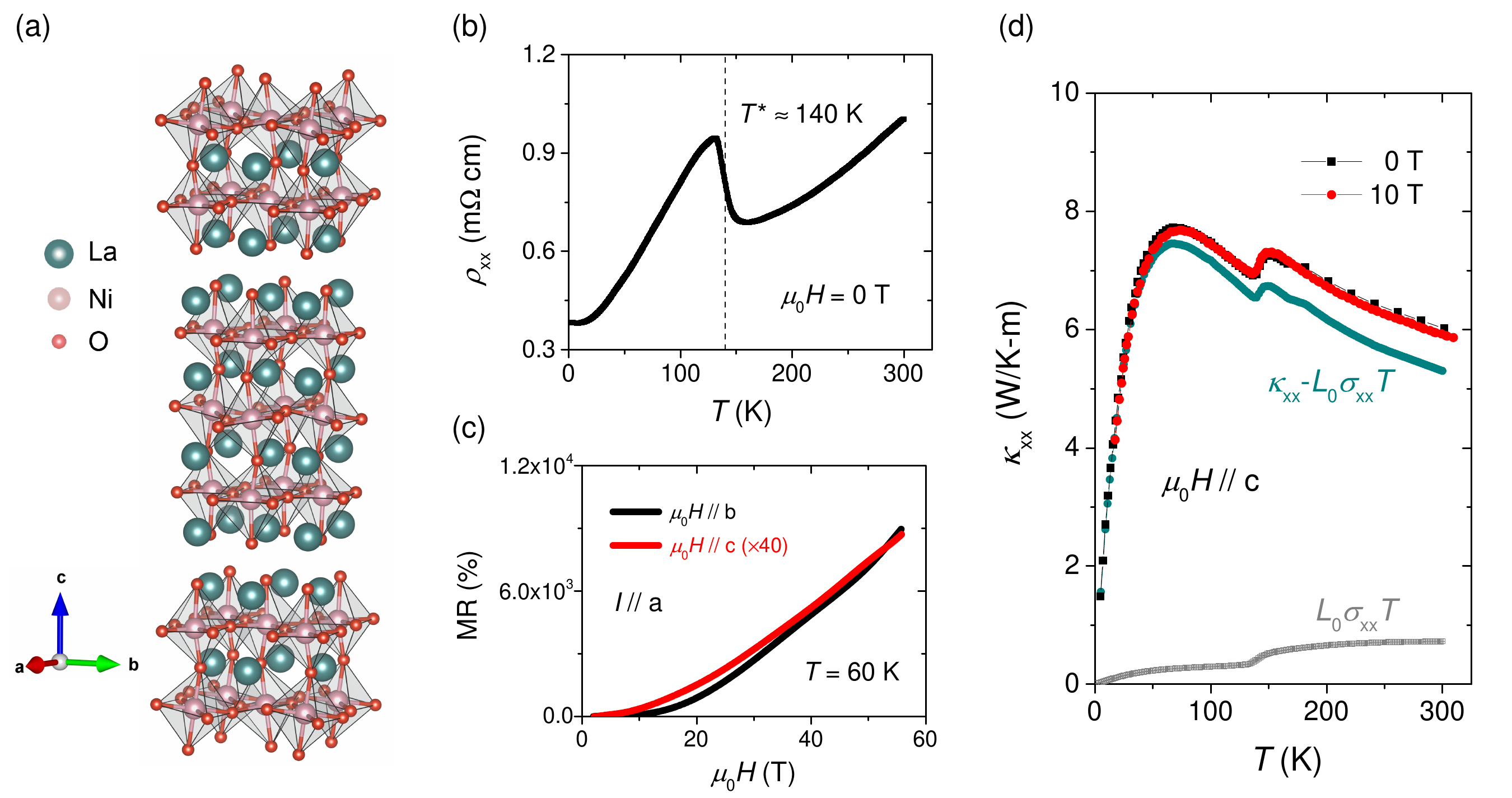} 
\caption{\textbf{Crystal structure and longitudinal transport properties of La$_4$Ni$_3$O$_{10}$ at ambient pressure.}
{(a)} Crystal structure of La$_4$Ni$_3$O$_{10}$ in the monoclinic $P2_1/a$ phase. La, Ni, and O atoms are indicated.
{(b)} Temperature dependence of the longitudinal resistivity $\rho_{xx}$ at zero field. It exhibits a clear anomaly near $T^*\approx140$~K, marking the density-wave transition. 
{(c)} Magnetoresistance (MR) measured under pulsed magnetic fields up to 55~T, where
$\mathrm{MR}(\%)=[\rho(B)-\rho(0)]/\rho(0)\times100\%$.
The MR for $\mu_0H\parallel b$ is much larger than that for $\mu_0H\parallel c$.
{(d)} Temperature dependence of the longitudinal thermal conductivity $\kappa_{xx}$ at 0~T and 10~T for $\mu_0H\parallel c$, together with the electronic contribution $L_0\sigma_{xx}T$. The nearly identical $\kappa_{xx}$ at both fields and the much smaller electronic term indicate that longitudinal heat transport is dominated by phonons.}
\label{fig:F1}
\end{figure*}

\begin{figure*}[ht]
\centering
\includegraphics[width=1\linewidth]{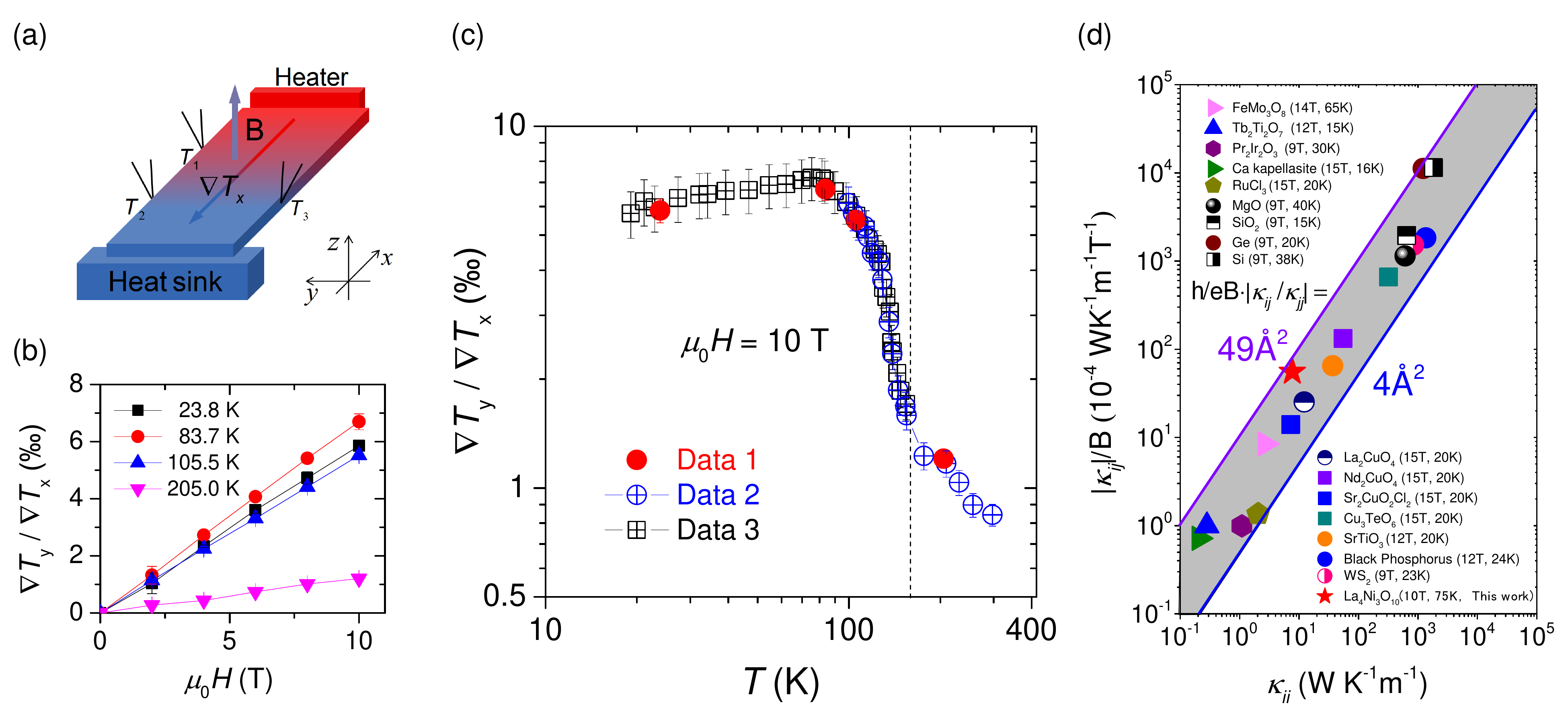} 
\caption{\textbf{Thermal Hall responses and its enhancement by density-wave order.}
{(a)} Schematic of the setup for simultaneous longitudinal and transverse thermal conductivity measurements.
{(b)} Thermal Hall angle $\nabla T_y/\nabla T_x$ vs magnetic field at selected temperatures, showing nearly linear dependence.
{(c)} Temperature dependence of $\nabla T_y/\nabla T_x$ at $\mu_0H=10$~T. The thermal Hall angle increases sharply below $T^*\approx140$~K and peaks around 70~K. Data from three different measurement methods (see Supplementary Materials~\cite{SM}) show good agreement.
{(d)} Comparison of the field-normalized transverse thermal conductivity, $|\kappa_{ij}|/B$, with the longitudinal thermal conductivity $\kappa_{ii}$ in La$_4$Ni$_3$O$_{10}$ and representative phonon thermal Hall systems. 
}
\label{fig:F2}
\end{figure*}

\begin{figure}[ht]
\centering
\includegraphics[width=1\linewidth]{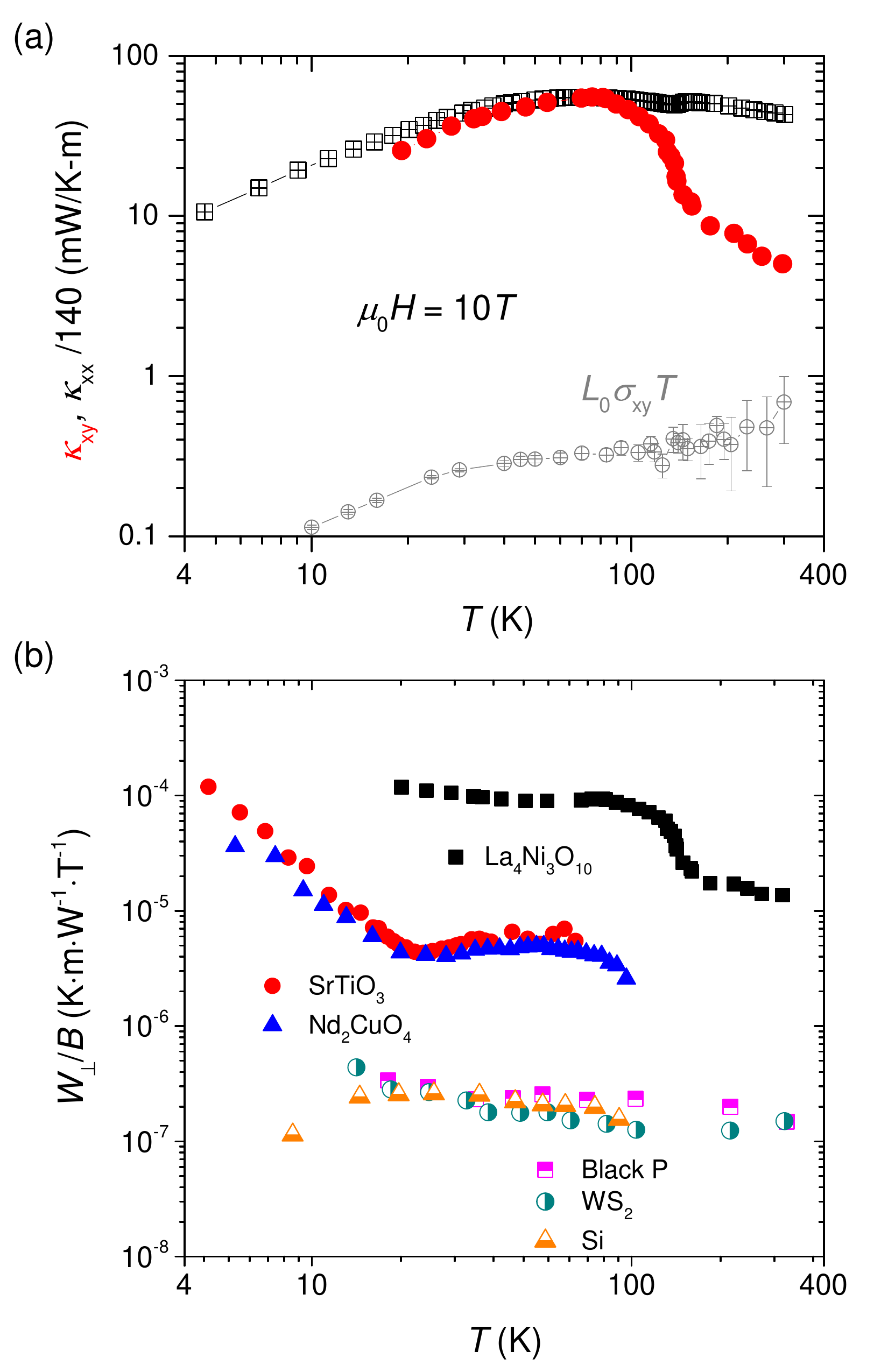} 
\caption{\textbf{Phonon thermal Hall conductivity and two thermal Hall resistivity plateaus.}
{(a)} Temperature dependence of $\kappa_{xy}$. The measured $\kappa_{xy}$ exceeds by far the electronic contribution (grey data), ruling out charge-carrier origin. Its peak coincides with that of $\kappa_{xx}$ (black data), a universal feature of the phonon thermal Hall effect. Together with linear field dependence and negligible field-dependent thermal conductivity (magnons contribute little), this establishes a phononic origin.
{(b)} Temperature dependence of the field-normalized thermal Hall resistivity, $W_\perp/B=(\nabla_yT/J_q^x)/B$, in La$_4$Ni$_3$O$_{10}$ and comparison with other solids\cite{Guo2026}. The value of La$_4$Ni$_3$O$_{10}$ pronounced change below $T^*$, indicates that the density-wave order strongly modifies transverse phonon transport. Two plateaus of thermal Hall resistivity can be clearly seen.
}
\label{fig:F3}
\end{figure}

\begin{figure*}[ht]
\centering
\includegraphics[width=1\linewidth]{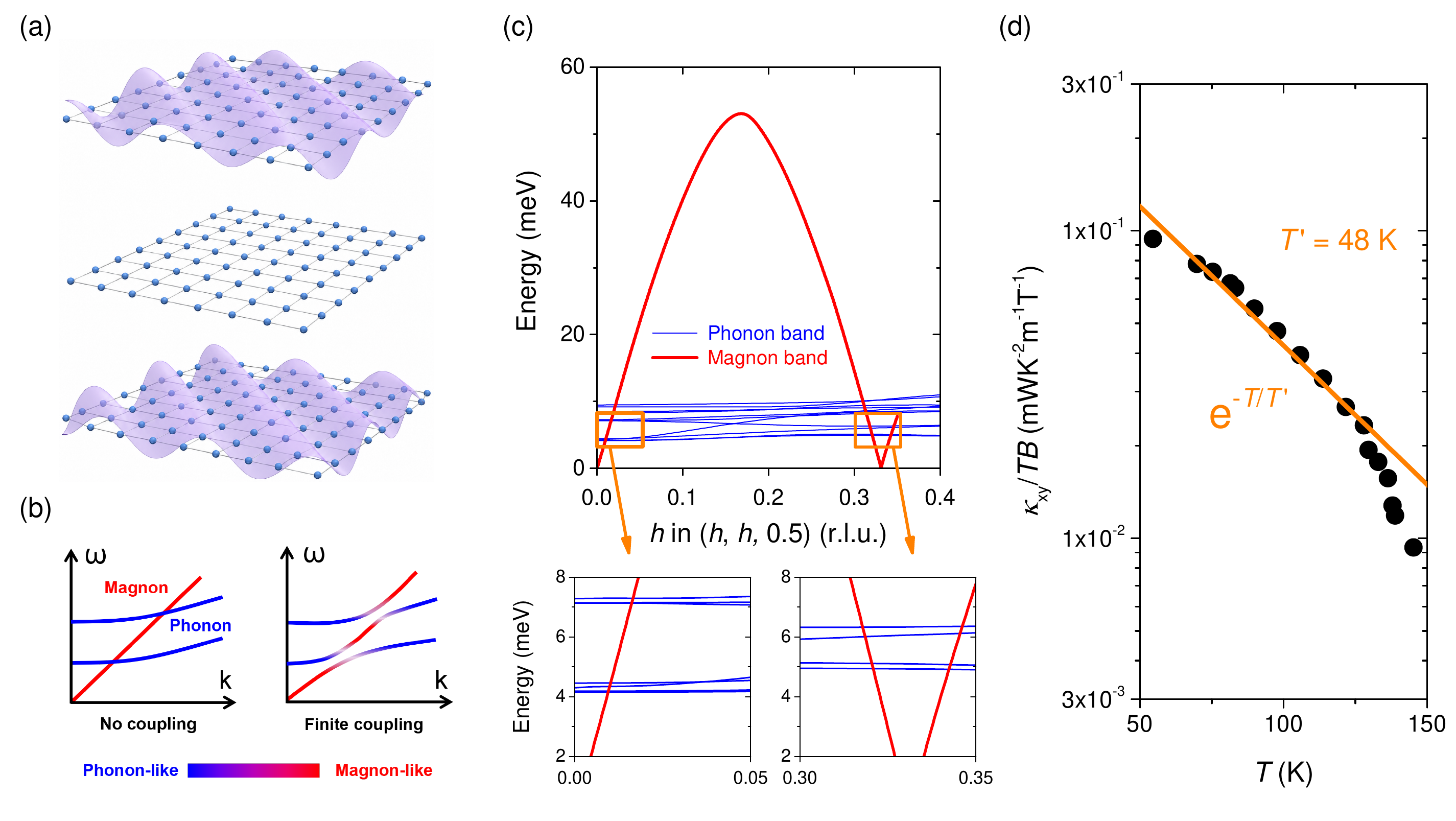} 
\caption{\textbf{Density wave and magnon-phonon hybridization.}
{(a)} Model of the SDW\cite{zhang2020intertwined}. The SDW is out of phase in the two outer NiO$_2$ layers and is almost absent in the inner layer.
{(b)} Schematic diagram of the magnon-phonon band hybridization. Without coupling, phonon and magnon bands are independent (left). Finite spin-lattice coupling induces their hybridization (right), imparting magnetic-field-sensitive properties (e.g., Berry curvature) to phonons, thereby enabling them to respond to a magnetic field.
{(c)} Calculated phonon bands (blue) \cite{Jia2026PRX} and modulated magnon dispersion (red)\cite{Chen2026SDW,Chan2026spin} of La$_4$Ni$_3$O$_{10}$ along the $(h,h,0.5)$(r.l.u.) momentum direction. Two sets of crossings (each set containing multiple crossing points) occur in the low-energy region near $h = 0$ and $h = 0.33$, respectively, as seen in the lower panels. The two sets of crossings span energy windows of approximately 3.2~meV and 1.4~meV, respectively.
{(d)} A semi-log plot of $\kappa_{xy}/(TB)$ as a function of temperature in La$_4$Ni$_3$O$_{10}$. Above the peak of $\kappa_{xy}$ (70 K) and below the DW transition (140 K), $\kappa_{xy}/(TB)$ approximately follows $\exp(-T/T')$, with $T'=48$~K, corresponding to an energy scale of $\sim4.1$~meV.
}
\label{fig:F4}
\end{figure*}

The thermal Hall effect---a transverse thermal response generated by a longitudinal heat current in the presence of a magnetic field---has emerged as a powerful probe of elementary excitations in quantum materials. In conventional conductors, the effect originates from the Lorentz deflection of charge carriers. Remarkably, finite thermal Hall signals have also been observed in a broad range of electrically insulating systems lacking mobile electrons\cite{Strohm2005,Onose2008,Onose2010,Hirsch2015,Watanabe2016,Ideue2017,Sugii2017,Li2017,Kasahara2018,Grissonnanche2019,Li2020,Grissonnanche2020,Boulanger2020,Akazawa2020,Sim2021,Chen2022,Uehara2022,Jiang2022,Bruin2022,Li2023,Chen2024,Chen2024-2,Ataei2024,meng2024thermodynamic,Li2025,Xiang2026}. In the parent insulating phase of cuprates, a sizable thermal Hall effect has been reported and attributed to chiral phonons and their coupling to magnetic degrees of freedom\cite{Grissonnanche2019,Grissonnanche2020,Boulanger2020,Chen2024}. In contrast, thermal Hall transport in nickelates remains largely unexplored. Among them, the trilayer compound La$_4$Ni$_3$O$_{10}$ is particularly attractive because it hosts a pronounced density-wave (DW) transition and can be synthesized as high-quality single crystals\cite{shi2025absence,zhu2024superconductivity,li2024signature}. These characteristics make it an ideal platform for investigating how DW order influences phonon dynamics through charge--lattice and spin--lattice coupling.

In this work, we uncover a pronounced phonon thermal Hall effect in the trilayer nickelate La$_4$Ni$_3$O$_{10}$ that is strongly enhanced by density-wave (DW) order. First, the thermal Hall angle, $\nabla T_y/\nabla T_x$, reaches a maximum value of nearly 7\textperthousand{} around 70 K---approximately twice that reported in cuprates\cite{Grissonnanche2019} and three times that observed in SrTiO$_3$\cite{Li2020}. The measured signal substantially exceeds the electronic contribution estimated from the Wiedemann--Franz law, while the negligible field dependence of the longitudinal thermal conductivity indicates only a minor magnon contribution to heat transport, pointing to a predominantly phononic origin. Further evidence comes from the close correspondence between the peaks in $\kappa_{xy}$ and $\kappa_{xx}$, a characteristic hallmark of phonon-mediated thermal Hall transport\cite{Li2020,Chen2022,Li2023,behnia2025phonon}. Second, the thermal Hall response is dramatically enhanced below the DW transition: the thermal Hall angle increases from approximately 1.5\textperthousand{} at 160 K to nearly 6\textperthousand{} at 100 K, signaling a strong coupling between the DW order and lattice excitations. Finally, we propose that magnon--phonon hybridization, arising from spin--lattice coupling associated with SDW correlations, serves as the primary mechanism underlying the enhanced thermal Hall effect. We further argue that this magnon--phonon hybridization, particularly involving low-energy optical phonon branches, may play a role in the suppression of the spin-density-wave order under pressure.

Fig.~\ref{fig:F1}(a) shows the ambient-pressure crystal structure of La$_4$Ni$_3$O$_{10}$ in the monoclinic $P2_1/a$ phase, consisting of trilayer NiO$_2$ planes separated by LaO rock-salt layers\cite{zhu2024superconductivity}. Fig.~\ref{fig:F1}(b) presents the zero-field longitudinal resistivity $\rho_{xx}(T)$. A pronounced inflection appears near the density-wave transition temperature $T^*\approx140$~K. Below $T^*$, $\rho_{xx}(T)$ exhibits a clear upturn, indicative of a partial gap opening\cite{escudero2010point,kesharpu2019temperature}, before recovering metallic behavior at lower temperatures. Fig.~\ref{fig:F1}(c) displays the magnetoresistance measured up to 55~T at 60~K for magnetic fields applied along the out-of-plane and in-plane directions. A pronounced anisotropy is observed: the magnetoresistance for $\mu_0H \parallel b$ is nearly 40 times larger than that for $\mu_0H \parallel c$. This behavior contrasts sharply with the conventional orbital magnetoresistance expected for a quasi-two-dimensional metal\cite{Liu2026anisotropy,Nie2022anisotropy}, where a stronger response is typically anticipated for out-of-plane fields. Combined with the density-wave transition at $T^{*}\approx140$~K, these results suggest that the magnetoresistance is governed predominantly by Zeeman coupling to an underlying spin/charge-ordered state, consistent with the anisotropic spin interaction $J$ in this system\cite{Chen2026SDW}.

The density-wave transition is also clearly reflected in the longitudinal thermal transport. As shown in Fig.~\ref{fig:F1}(d), $\kappa_{xx}$ exhibits a distinct anomaly around $T^*$, with the data measured at 0~T and 10~T nearly overlapping. This weak field dependence indicates that magnetic-field-sensitive excitations, such as magnons, hardly affect the longitudinal heat transport. Moreover, the electronic thermal conductivity estimated from the Wiedemann--Franz law, $\kappa_{xx}^{e}=L_0\sigma_{xx}T$, accounts for only a small fraction of the measured total $\kappa_{xx}$. These results demonstrate that longitudinal heat transport in La$_4$Ni$_3$O$_{10}$ is predominantly phononic. The change in $\kappa_{xx}$ near $T^*$ is larger than the value which can be explained by the modification of electronic transport alone, implying that the density-wave order directly influences lattice vibrations (phonons).

We next investigated the transverse thermal transport properties. Fig.~\ref{fig:F2}(a) shows the experimental setup for simultaneous longitudinal and transverse thermal conductivity measurements. Fig.~\ref{fig:F2}(b) shows the magnetic field dependence of the thermal Hall angle $\nabla T_y/\nabla T_x$ at several fixed temperatures. It exhibits a well-defined linear dependence. Fig.~\ref{fig:F2}(c) presents the temperature dependence of $\nabla T_y/\nabla T_x$ at $\mu_0H=10$~T. Data from three independent measurement methods agree well (see Supplementary Materials~\cite{SM}). The thermal Hall response is sharply enhanced below $T^*$: the thermal Hall angle increases from 1.5\textperthousand{} at 160~K to 6\textperthousand{} at 100~K (fourfold), reaching a maximum of 7\textperthousand{} near 70~K. This value is twice that in cuprates\cite{Grissonnanche2019}, three times that in SrTiO$_3$\cite{Li2020}, and close to the upper bound of the phonon thermal Hall angle [Fig.~\ref{fig:F2}(d)]\cite{Li2023}. This pronounced enhancement below $T^*$ indicates strong coupling between density-wave order and phonons. 

Fig.~\ref{fig:F3}(a) compares the measured transverse thermal conductivity $\kappa_{xy}$ with the electronic contribution estimated from the Wiedemann--Franz law. The measured $\kappa_{xy}$ exceeds by far the electronic contribution: it is roughly an order of magnitude larger at high temperatures and more than two orders of magnitude below 100~K, indicating that electronic carriers contribute negligibly to transverse heat transport. More importantly, the estimated electronic term shows only a weak anomaly near $T^*$ and exhibits a temperature dependence opposite to that of the measured $\kappa_{xy}$. This stark contrast unambiguously demonstrates that the pronounced enhancement of $\kappa_{xy}$ below $T^*$ does not originate from conventional charge carriers, but rather from other quasiparticles. Together with the coincidence of the $\kappa_{xy}$ and $\kappa_{xx}$ peaks (a universal feature of the phonon thermal Hall effect\cite{Li2020,Chen2022,Li2023,behnia2025phonon}), and the negligible field-dependent thermal conductivity (indicating that magnons contribute little), these observations establish a predominantly phononic origin. 

In this system (as well as in cuprates), despite the presence of a large phonon thermal Hall effect and a finite electronic contribution, there is no phonon-drag thermal Hall effect---in contrast to SrTiO$_3$ and graphite---because the electronic mobility remains low compared to those in graphite and dilute SrTiO$_3$\cite{Jiang2022,Xiang2026}. A phonon-drag thermal Hall effect requires a large electronic Hall angle, which is not realized in metallic nickelates or cuprates. 

Fig.~\ref{fig:F3}(b) shows the temperature dependence of the field-normalized thermal Hall resistivity, $W_{\perp}/B=\kappa_{xy}/(\kappa_{xx}^{2}B)=(\nabla_yT/J_q^x)/B$, in La$_4$Ni$_3$O$_{10}$ and in other solids\cite{Guo2026}. The transition-induced enhancement is more clearly resolved by the observation of two distinct thermal Hall resistivity plateaus in La$_4$Ni$_3$O$_{10}$. This behavior suggests that both below and above $T^*$, a transverse Lorentz-like force, set by the product of the magnetic field and the heat flux, is counterbalanced by an entropic force.  But for some reason, the Lorentz-like force is an order of magnitude larger in the low-temperature phase. Meanwhile, the magnitude of $W_{\perp}/B$ in La$_4$Ni$_3$O$_{10}$ is comparable to that in Nd$_2$CuO$_4$ and SrTiO$_3$, but nearly two orders of magnitude larger than in clean solids such as Si, black phosphorus, and WS$_2$\cite{Guo2026}. This contrast between insulators hosting ballistic phonons and others indicate that the coupling between magnetic field and heat flux is intensified when acoustic phonons are not the only relevant quasi-particles.

In La$_4$Ni$_3$O$_{10}$, the density-wave transition involves an incommensurate out-of-phase SDW order [Fig.~\ref{fig:F4}(a)] and an in-phase CDW order\cite{zhang2020intertwined,chen2024trilayer}.  Resonant inelastic X-ray scattering measurements, when interpreted with linear spin-wave theory, reveal that the spin-density wave order gives rise to well-defined magnon bands, including both acoustic and optical branches\cite{Chen2026SDW,Chan2026spin}.  

It has recently been proposed that a possible source of the thermal Hall effect in NiPS$_3$ is the hybridization of acoustic phonons and magnons\cite{meng2024thermodynamic}. An alternative hybridization picture between phonons and magnons involves low-energy optical phonons and acoustic magnons, as shown in Fig.~\ref{fig:F4}(b). As shown in Fig.~\ref{fig:F4}(c), the calculated phonon bands\cite{Jia2026PRX} and modulated magnon dispersion of La$_4$Ni$_3$O$_{10}$\cite{Chen2026SDW,Chan2026spin} along the $(h,h,0.5)$ (r.l.u.) momentum direction cross near $h = 0$ and $h = 0.33$, with a crossing energy span of $\sim$3.2~meV and $\sim$1.4~meV respectively. Yang et al.\cite{Yang2020universal} argued that, independent of microscopic details, an intrinsic thermal Hall effect follows the simple form $\kappa_{xy}/T\propto\exp(-T/T')$, where $T'$ represents the characteristic temperature corresponding to the energy window over which the Berry curvature is finite and constant. A fit to the temperature dependence of $\kappa_{xy}$ [Fig.~\ref{fig:F4}(d)] yields $T'=48$~K ($\sim4.1$~meV), which is comparable to the above-mentioned 3.2~meV crossing energy span. The remaining small discrepancy and the slight imperfection of the fit may originate from the contribution of the crossing near $h = 0.33$. Furthermore, the average energy scale of the hybridized phonon band is about 6 meV, which recalls the thermal Hall peak at 70 K.

From cuprates to iron-based superconductors, and now to nickelates, the interplay among magnetism, the lattice, and superconductivity has evolved markedly. Across these diverse systems, the suppression of static magnetic order---often achieved by doping or pressure---appears to be a key prerequisite for the emergence of superconductivity. In La$_4$Ni$_3$O$_{10}$ at ambient pressure, the system already hosts an incommensurate spin-density-wave order below $T^*$\cite{khasanov2026pressure}, yet no long-range commensurate antiferromagnetic order is observed. Intriguingly, our results reveal that optical phonons are dynamically coupled to spin excitations in this material. Such coupling can soften the optical phonon modes, a scenario that is expected to strengthen the spin--lattice interaction. Under applied pressure, this enhanced coupling may help destabilize the spin-density-wave state, thereby facilitating the onset of superconductivity.

Notably, the interlayer apical oxygen in La$_4$Ni$_3$O$_{10}$ is highly dynamic and may play an important role in the coupled charge--lattice physics of the system. It is prone to displacement, may host holes in its $p_z$ orbitals, and forms inequivalent bonds with the inner and outer Ni ions, leading to strong lattice anharmonicity and enhanced charge--lattice coupling\cite{sun2023signatures}. An alternative mechanism for the enhanced thermal Hall response is the circular lattice motion related to charge--lattice coupling. In the trilayer Ni--O network, polar bonds between the apical oxygen ($2p$) orbitals and neighboring Ni ($3d_{z^2}$) orbitals generate local electric dipoles $P$. Under an out-of-plane magnetic field $B$, acoustic lattice vibrations can induce circular motion of the ions (see Supplementary Materials~\cite{SM}), generating phonon angular momentum that contributes to the phonon thermal Hall effect\cite{Zhang2014,Li2025}.

In La$_4$Ni$_3$O$_{10}$, we observe an enhanced phonon thermal Hall effect tied to the density-wave transition. We propose that spin-lattice coupling (magnon-phonon hybridization) from SDW order drives both the thermal Hall enhancement and the suppression of long-range antiferromagnetism. Thus, La$_4$Ni$_3$O$_{10}$ exemplifies intertwined charge, spin, and lattice orders, where phonons dressed by electronic order play an active role in the normal state of superconducting nickelates.

\vspace*{4mm}

\textit{Acknowledgments}—We are grateful to Kamran Behnia, Zhengyu Weng, Daoxin Yao, Qisi Wang and Meng Wang for stimulating discussions. This work was supported by the National Key Research and Development Program of China (Grants No. 2023YFA1609600, 2024YFA1611200, and 2022YFA1403500), the National Science Foundation of China (Grants No. 12304065, 51821005, 12004123, 51861135104, and 11574097), the Fundamental Research Funds for the Central Universities (Grant No. 2019kfyXMBZ071), and the Hubei Provincial Natural Science Foundation (Grant No. 2025AFA072).

\vspace*{3mm}

Qiaochao Xiang and Enkang Zhang contributed equally to this work.

\vspace*{3mm}

\noindent
* \verb|lixiaokang@hust.edu.cn|\\
* \verb|zhaoj@fudan.edu.cn|\\
* \verb|zhanggm@shanghaitech.edu.cn|\\
* \verb|Liangli44@hust.edu.cn|\\
* \verb|zengwei.zhu@hust.edu.cn|\\

\bibliography{main}

\clearpage
\renewcommand{\thesection}{S\arabic{section}}
\renewcommand{\thetable}{S\arabic{table}}
\renewcommand{\thefigure}{S\arabic{figure}}
\renewcommand{\theequation}{S\arabic{equation}}
\setcounter{section}{0}
\setcounter{figure}{0}
\setcounter{table}{0}
\setcounter{equation}{0}

{\large\bf Supplementary Material for ``Density-wave order enhances the phonon thermal Hall effect in a trilayer nickelate''}
{\large\bf by Q. Xiang et al.}

\setcounter{figure}{0}

\section{Sample and Methods}

The precursor powder of La$_4$Ni$_3$O$_{10}$ was prepared using a conventional solid-state reaction method. After the powder was pressed into a cylindrical rod under high pressure and sintered, single crystal growth was carried out in a vertical optical-image floating-zone furnace (Model HKZ, SciDre). For more details, see Ref.\cite{zhu2024superconductivity}. The sample used in this work had approximate dimensions of 3.0 mm (length) $\times$ 1.7 mm (width) $\times$ 1.2 mm (thickness).

Steady-field transport experiments were performed in a commercial measurement system (Quantum Design PPMS) with a stable high-vacuum sample chamber. Voltages were monitored using DC nanovoltmeters (Keithley 2182A), and electric current was supplied by a current source (Keithley 6221). A one-heater, four-thermocouple (type E) method was employed to simultaneously measure the longitudinal and transverse thermal gradients. The thermal gradient in the sample was generated by a 9.0 k$\Omega$ chip resistor powered by a current source (Keithley 6221). The thermocouples, heat sink, and heater were directly attached to the sample using silver paste for all contacts. Heat current was applied along the crystallographic $a$ axis, while the transverse thermal gradient was measured along the in-plane direction perpendicular to the heat current. The magnetic field was applied along the out-of-plane direction, $\mu_0H\parallel c$, unless otherwise specified. The field-antisymmetric component of the transverse temperature difference was extracted as $\Delta T_y^{\rm asym}(H)=[\Delta T_y(+H)-\Delta T_y(-H)]/2$, which removes the contribution from longitudinal thermal gradients caused by contact misalignment.

To measure the thermal Hall angle efficiently, we employed three complementary methods.
In the first method, at a fixed temperature, the magnetic field was swept from $+10$\,T to $-10$\,T and back to $+10$\,T in steps of 2\,T, and the data were antisymmetrized; this confirmed the linear field dependence of the signal. Based on this linearity, the second method only required measuring the transverse temperature difference at $+10$\,T and $-10$\,T for each temperature, significantly improving efficiency.
In the third method, we performed a temperature sweep at $+10$\,T, then reversed the field to $-10$\,T and repeated the same temperature sweep. The agreement among the three methods confirms the reproducibility of the thermal Hall signal.

Magnetoresistance (MR) measured under pulsed magnetic fields up to 55~T was measured in a pulsed magnetic field equipment at the Wuhan National High Magnetic Field Center (WHMFC).

\section{Magnetoresistance, Hall resistivity, and magnetothermal conductivity}

\begin{figure*}[ht]
\centering
\includegraphics[width=1\linewidth]{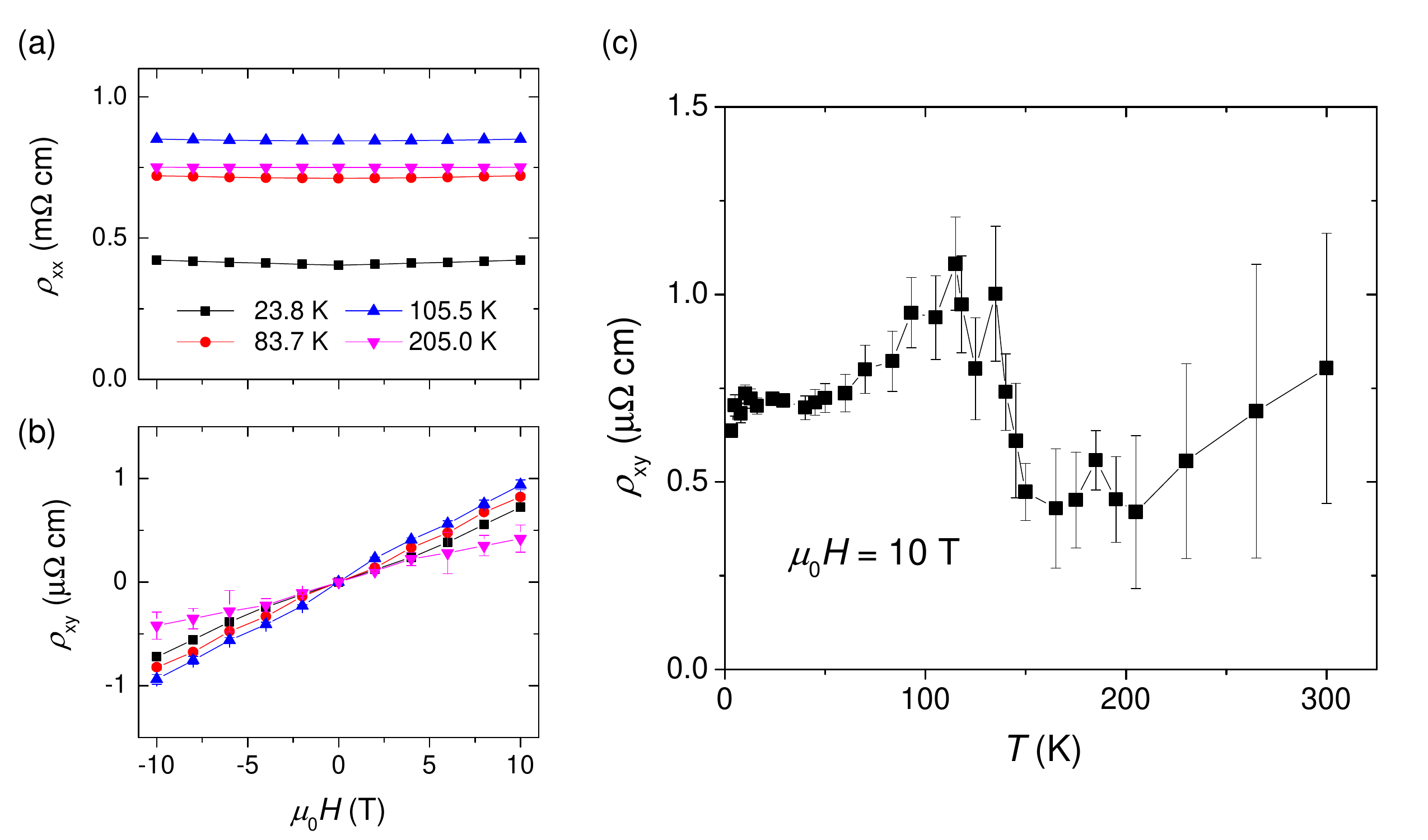} 
\caption{\textbf{Longitudinal and Hall resistivity.}
{(a)} Field dependence of the longitudinal resistivity $\rho_{xx}$ at four selected temperatures.
{(b)} Field dependence of the Hall resistivity $\rho_{xy}$ at the same temperatures.
{(c)} Temperature dependence of $\rho_{xy}$ at $\mu_0H=10$~T.}
\label{fig:S1}
\end{figure*}

The magnetic-field dependence of the longitudinal resistivity and Hall resistivity was measured at four representative temperatures, as shown in Figs.~\ref{fig:S1}(a) and (b), respectively. The sample exhibits weak magnetoresistance below 10 T. The Hall resistivity remains nearly linear in field throughout the measured temperature range, indicating hole-dominated transport. The temperature dependence of the Hall resistivity at 10\,T is summarized in Fig.~\ref{fig:S1}(c).

\begin{figure*}[ht]
\centering
\includegraphics[width=0.5\linewidth]{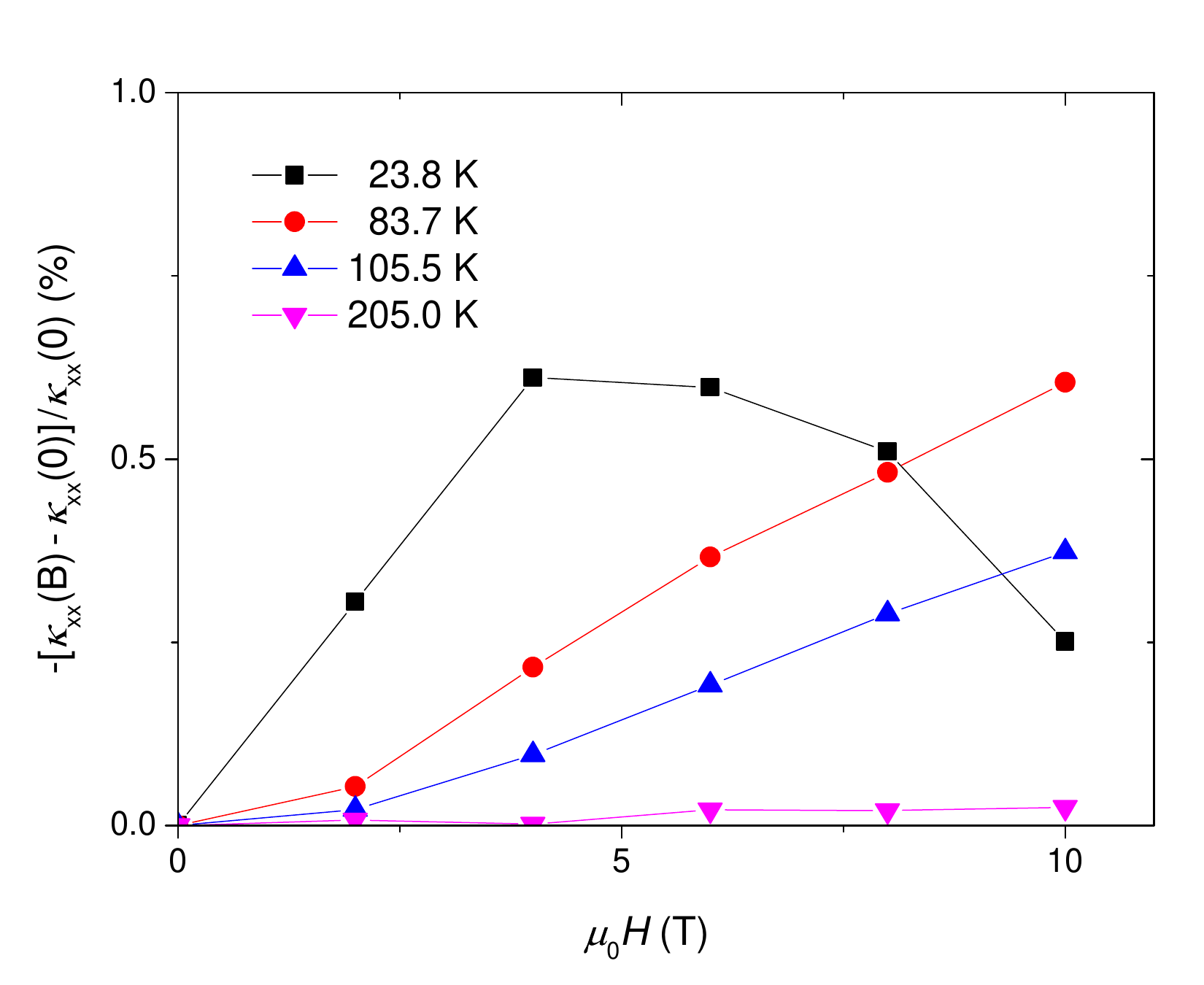}
\caption{\textbf{Magnetothermal response of longitudinal thermal conductivity.} 
Field dependence of the percentage change $-\left[\kappa_{xx}(B)-\kappa_{xx}(0)\right]/\kappa_{xx}(0)\times100\%$ at four temperatures.
The change is below 1\% for all fields, showing that $\kappa_{xx}$ is insensitive to the magnetic field.}
\label{fig:S2}
\end{figure*}

We also examined the magnetic response of the longitudinal thermal conductivity. Fig.~\ref{fig:S2} shows the percentage change $-\left[\kappa_{xx}(B)-\kappa_{xx}(0)\right]/\kappa_{xx}(0)\times100\%$ as a function of magnetic field at four representative temperatures.
The variation remains below 1\% up to 10\,T across the entire measured range, demonstrating that $\kappa_{xx}$ is essentially insensitive to the applied field. This negligible magnetothermal response indicates that magnons contribute minimally to the longitudinal heat transport.

\section{Negligible phonon drag contribution}
\begin{figure*}[ht]
\centering
\includegraphics[width=1\linewidth]{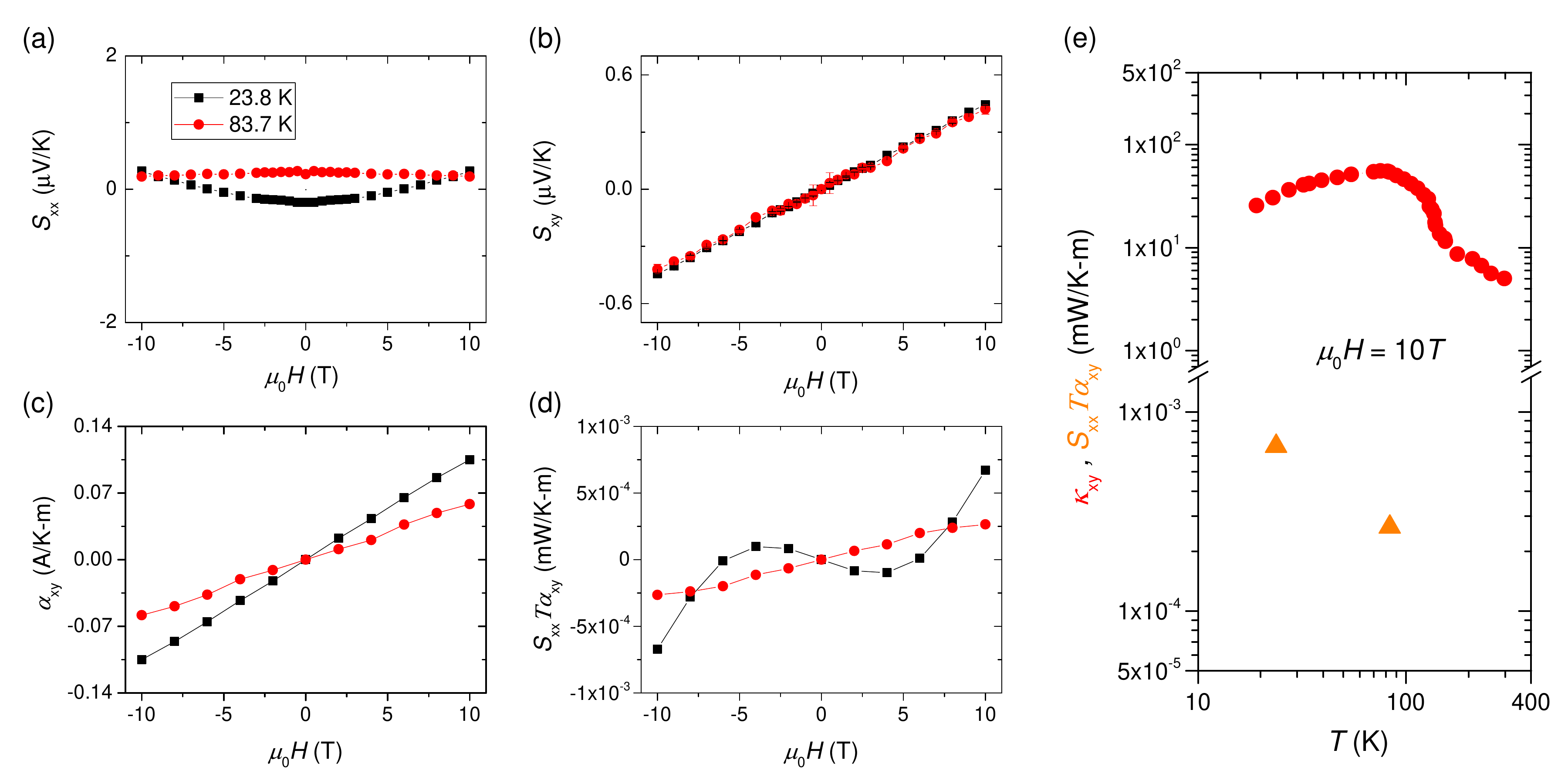} 
\caption{\textbf{Thermoelectric properties and phonon drag.}
{(a)} Field dependence of the Seebeck coefficient at 23.8~K and 83.7~K.
{(b)} Field dependence of the Nernst coefficient at the same temperatures.
{(c)} Nernst conductivity $\alpha_{xy}$ obtained from $\alpha_{xy}=(\rho_{xx}S_{xy}-\rho_{xy}S_{xx})/\rho_{xx}^2$.
{(d)} Upper bound of the phonon-drag contribution $\kappa_{xy}^{\mathrm{drag}}(\mathrm{max})=S_{xx}T\alpha_{xy}$ at 23.8~K and 83.7~K.
{(e)} Comparison of the measured $\kappa_{xy}$ at 10~T with the phonon-drag estimate $S_{xx}T\alpha_{xy}$, which is four orders of magnitude smaller.}
\label{fig:S3}
\end{figure*}

Figs.~\ref{fig:S3}(a) and (b) show the field dependence of the Seebeck and Nernst coefficients at 23.8\,K and 83.7\,K.
As in the electrical transport, the longitudinal thermoelectric response depends only weakly on field, while the transverse response is nearly linear in field.
The Nernst conductivity $\alpha_{xy}$ [Fig.~\ref{fig:S3}(c)] is obtained via
\begin{equation}\label{eq:nernst} 
\alpha_{xy}=\frac{\rho_{xx}S_{xy}-\rho_{xy}S_{xx}}{\rho_{xx}^2},
\end{equation}
where $\rho_{xx}$, $\rho_{xy}$, $S_{xx}$, $S_{xy}$ denote the resistivity, Hall resistivity, Seebeck coefficient, and Nernst coefficient, respectively.

The phonon-drag contribution to $\kappa_{xy}$ is $\kappa_{xy}^{\rm drag}=S_{\rm drag}T\alpha_{xy}$.
To estimate its upper bound, we assume the extreme case that the entire measured Seebeck coefficient is of phonon-drag origin, i.e.\ $S_{\rm drag}=S_{xx}$.
The maximum possible phonon-drag contribution is then
\begin{equation}\label{eq:dragmax}
\kappa_{xy}^{\rm drag}({\rm max})=S_{xx}T\alpha_{xy}.
\end{equation}
As shown in Figs.~\ref{fig:S3}(d) and (e), this value is nearly four orders of magnitude smaller than the measured $\kappa_{xy}$, ruling out phonon drag as the origin of the thermal Hall signal.

\section{Charge--lattice coupling from apical oxygen dynamics}
\begin{figure*}[ht]
\centering
\includegraphics[width=0.75\linewidth]{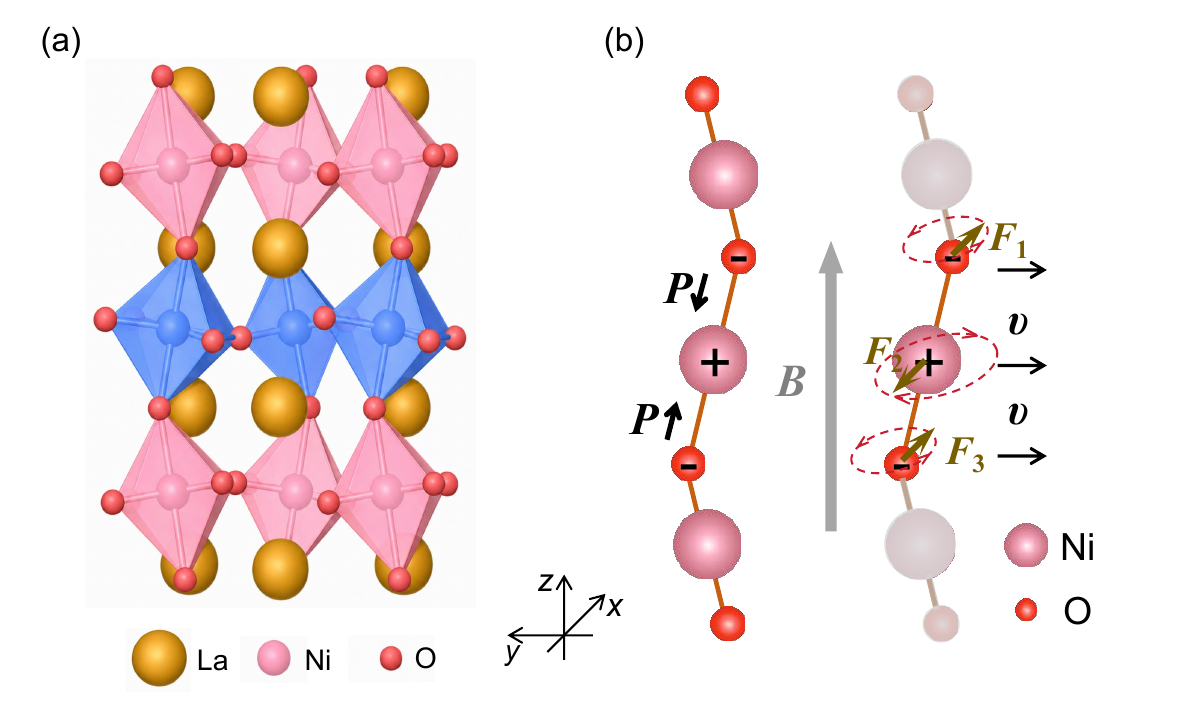} 
\caption{\textbf{Charge--lattice coupling through apical oxygen.}
{(a)} Trilayer network of NiO$_6$ octahedra in the $P2_1/a$ phase of La$_4$Ni$_3$O$_{10}$, with each layer drawn in a single color.
{(b)} Schematic illustration of the charge--lattice coupling mechanism. Hybridization between apical O~$2p$ and Ni~$3d_{z^2}$ orbitals creates local electric dipoles~$P$. Lattice vibrations displace the charge centers, producing oscillating dipoles. An out-of-plane magnetic field exerts Lorentz forces $\mathbf{F}=q\mathbf{v}\times\mathbf{B}$ on the moving ions, driving a circular motion that generates a phonon angular momentum, thereby contributing to the transverse phonon response.}
\label{fig:S4}
\end{figure*}

The trilayer structure of La$_4$Ni$_3$O$_{10}$ contains highly flexible apical oxygen atoms that connect adjacent NiO$_2$ layers, as illustrated in Fig.~\ref{fig:S4}(a).
The inequivalent bonding environments at the inner and outer Ni sites lead to asymmetric Ni--O bond lengths and angles, giving rise to strong lattice anharmonicity and enhanced charge--lattice coupling\cite{sun2023signatures}.

A possible microscopic picture is shown in Fig.~\ref{fig:S4}(b). Hybridization between apical O~$2p$ and neighboring Ni~$3d_{z^2}$ orbitals forms polar Ni--O bonds with local electric dipoles~$P$. During lattice vibrations, the displacement of these charge centers produces oscillating dipoles. Under an out-of-plane magnetic field, the moving ions experience Lorentz forces, inducing a circulating ionic motion and a phonon angular momentum\cite{Zhang2014,Li2025}.
Such field-coupled lattice dynamics could provide an additional contribution to the transverse phonon response.

\end{document}